\documentstyle[12pt]{article}
\input epsf
\newcommand{\beq}{\begin{eqnarray}}   
\newcommand{\eeq}{\end{eqnarray}}

\newcommand{\gsim}{\lower.7ex\hbox{$
\;\stackrel{\textstyle>}{\sim}\;$}}
\newcommand{\lsim}{\lower.7ex\hbox{$
\;\stackrel{\textstyle<}{\sim}\;$}}

\input epsf
\newcommand{\grpicture}[1]
{
    \begin{center}
        \epsfxsize=300pt
        \epsfysize=200pt
        \vspace{-5mm}
        \parbox{\epsfxsize}{\epsffile{#1.eps}}
        \vspace{5mm}
    \end{center}
}

\begin{document}
\begin{titlepage}

\begin{flushright}
ITEP-TH-61/97 \\
Saclay/SPhT/T97-130

\end{flushright}
\vspace*{3cm}

\begin{center}
{\Large \bf   BPS Domain Walls   in Supersymmetric $QCD$ : Higher
Unitary Groups}
\vspace{2cm}

{\Large  A.V. Smilga} \\

\vspace{0.8cm}

{\it ITEP, B. Cheremushkinskaya 25, Moscow 117218, Russia}\\
 and \\
{\it Service de Physique Th\'eorique de Saclay, 91191 Gif--Sur--Yvette,
France} 

\end{center}

\vspace*{2cm}

\begin{abstract}
We consider the effective lagrangian 
describing the ${\cal N} = 1$ supersymmetric $QCD$ with  $SU(N_c)$ gauge 
group involving $N_f = N_c - 1$ pairs of chiral matter multiplets in 
fundamental and antifundamental color
representations. For this theory in the framework of the effective lagrangian
approach, we solve BPS equations for the domain walls
interpolating between different vacua. The equations always have a unique 
solution for the walls interpolating between
the chirally asymmetric and a chirally symmetric vacua. For the walls 
interpolating between different  chirally asymmetric vacua, the
equations admit two different solutions which exist in a limited range of 
the mass of the matter fields $m < m_* = \kappa \Lambda_{SQCD}$ where the
 parameter
$\kappa$ depends on $N_c$. At $m = m_*$, two branches join together and,
at $m > m_*$, no BPS -- saturated complex domain walls exist.
\end{abstract}

\end{titlepage}

\section{Introduction}

Supersymmetric QCD is the theory involving  a gauge
 vector supermultiplet
$V$ and an even number of chiral matter supermultiplets $S_{i}$, $S'_i$ 
($i = 1,\ldots, N_f$ is the flavour index)
such that $S_{i}$ belonging to the fundamental representation of the gauge
group $SU(N_c)$ involve left quarks and their scalar superpartners 
 and $S'_{i}$ belonging to the antifundamental representation involve left 
antiquarks and their scalar superpartners.
 The lagrangian of the model reads
\beq
{\cal L} = \left( \frac{1}{4g^2} \mbox{Tr} \int d^2\theta \ W^2\ + \ {\rm H.c.}
\right)\ + \  \sum_i \left( \frac{1}{4}\int d^2\theta d^2\bar\theta\ 
\bar S_{i}  e^V S_{i}  \right. \nonumber \\
+ \left.  \frac{1}{4}\int d^2\theta d^2\bar\theta\ 
 S'_{i}  e^{-V} \bar S'_{i} \right)
 - \left( \frac{m_{ij}}{2} \int d^2\theta\  S'_i S_{j} 
+\mbox{H.c.}\right)\, ,
\label{LSQCD}
\eeq
 $m_{ij}$ is the  complex mass 
matrix; color and Lorentz indices are suppressed. 

The dynamics of this model attracted the attention of theorists since the
beginning of the eighties. In some respects, it is similar to the dynamics
of the standard (non--supersymmetric) QCD. If the number of flavours $N_f$ is 
not too large, it involves confinement, for instance.
The model (\ref{LSQCD}) has, however, many specific features which are due
to supersymmetry. Actually, this theory is in some sense much simpler than the
standard QCD --- the presence of the extra symmetry allows one to obtain many 
exact results. The results are especially strong for the extended 
${\cal N} = 2$ version of the theory \cite{SW}, but
 a lot of exact theorems can be derived also for ${\cal N} = 1$ theory
 \cite{brmog}.

The details of the dynamics depend on $N_f$ and on the matter mass matrix
$m_{ij}$. A common feature of all models (\ref{LSQCD}) is their nontrivial
vacuum structure associated with the spontaneous breaking of a discrete chiral
symmetry.
Like in the standard QCD, the axial $U_A(1)$ symmetry corresponding to
the chiral rotation of the gluino field and present in the tree--level
lagrangian (\ref{LSQCD}) is broken by anomaly down to $Z_{2N_c}$. This
discrete chiral symmetry can be further broken spontaneously down to $Z_2$
so that the chiral condensate $<{\rm Tr}\ \{ \lambda^\alpha \lambda_\alpha\}>$ 
($\lambda_\alpha$ is a two - component Weyl spinor describing the gluino
field) is formed. There are
$N_c$ different vacua with different phases of the condensate
  \beq
  <{\rm Tr}\ \lambda^2> \ =\ \Sigma e^{2\pi i k/N_c},\ \ \
  \ \ \  k = 0, \ldots, N_c-1
  \label{cond}
   \eeq
   It was noted recently \cite{Kovner} that on top of $N_c$ chirally
asymmetric vacua (\ref{cond}), also a chirally symmetric vacuum with
zero value of the condensate exists.

The presence of different degenerate physical vacua in the theory
implies the existence of domain walls --- static field configurations 
depending  only on one spatial coordinate ($z$) which interpolate between
one of the vacua at $ z = -\infty$ and another one at $z = \infty$ and
minimizing the energy functional. As was shown in \cite{Dvali}, in many
cases the energy density of these walls can be found exactly due to the
fact that the walls present the BPS--saturated states. 
\footnote{Such BPS--saturated walls were known earlier in
2--dimensional supersymmetric theories (they are just solitons there)
\cite{WO,Vafa} and were considered also in 4--dimensional (non--gauge) 
theories in stringy context \cite{Rey}. }.

One can show (see Refs.\cite{my,Chib} for detailed derivation and discussion)
that the energy density of a BPS--saturated wall in SQCD satisfies a relation
  \beq
   \label{eps}
   \epsilon \ =\ \frac {N_c}{8\pi^2} \left|<{\rm Tr}\ \lambda^2>_\infty
    \ -\ <{\rm Tr}\ \lambda^2>_{-\infty} \right|
    \eeq
where the subscript $\pm \infty$ marks the values of the gluino
condensate at spatial infinities. The RHS of Eq.(\ref{eps}) presents an 
absolute   lower bound for the energy of any field configuration interpolating
between different vacua.
    
    The relation (\ref{eps}) is valid {\it assuming} that the wall is
BPS--saturated. However, whether such a BPS--saturated domain wall
exists or not is a non--trivial dynamic question which can be answered
only in a specific study of a particular theory in interest. 

In Refs.\cite{my,SV,SVn} this question was studied in the simplest nontrivial
case $N_c = 2,\ N_f = 1$
in the framework of the effective low energy lagrangian due to Taylor,
Veneziano, and Yankielowicz \cite{TVY}. In that case, it
 is written in terms of the composite colorless chiral superfields
  \beq
  \label{norm}
\Phi^3 = \frac 3{32\pi^2} {\rm Tr}\ W^2, \ \ X^2 =\ 2 S'S
  \eeq
and  presents a Wess--Zumino model with the superpotential    
\beq
{\cal W} = \frac{2}{3} \Phi^3 \left[ \ln \frac{\Phi^3 X^2}
{\Lambda_{SQCD}^5} \ -\ 1 \right] - \frac{m}{2} X^2 
\label{WTVY2}
\eeq
The results are the following: 
\begin{enumerate}
\item For any value of the mass of the matter
fields $m$, there are domain walls interpolating between a chirally asymmetric
and the chirally symmetric vacua. They are BPS -- saturated. 
  \item There are also complex BPS solutions interpolating between different
chirally asymmetric vacua. But they exist only if the mass is small enough
$m \leq m_* = 4.67059\ldots \Lambda_{SQCD}$. When $m > m_*$, BPS walls are
absent.
  \item In a narrow range of masses $m_* < m \leq m_{**} \approx 4.83 
\Lambda_{SQCD}$, complex domain walls still exist, but they are not BPS 
saturated anymore. At $m > m_{**}$, there are no such walls whatsoever.
 \end{enumerate}
 All these results were obtained by solving numerically the first order 
BPS equations
   \beq
  \partial_z \phi \ =\ e^{i\delta} \partial \bar {\cal W} /\partial \bar 
\phi,  \ \ \ \ \      \partial_z \chi \ =\ e^{i\delta} 
\partial \bar {\cal W} /\partial \bar \chi\ 
  \label{BPS}
  \eeq
associated with the TVY lagrangian
\footnote{In the $SU(2)$ case, we should set $\delta = 0$ or $\delta = \pi$
depending on the ``direction'' of the wall.}
(see Refs.\cite{my,Chib} for details) and/or the equations of motion for the
 profile of the wall with the
 proper boundary conditions.
Technically, the easiest problem was to study the point 1 of the list above
\cite{my}: the phase of the
complex scalar fields $\phi$ and $\chi$ (the lowest components of the 
superfields $\Phi$ and $X$) does not change along such a wall  and can be
set to zero which simplifies
the equations considerably. Point 2 is a  bit more 
difficult \cite{SV}: one has to solve here essentially complex BPS equations
 to obtain essentially  complex  solutions.
Point 3 is still more difficult: we have to solve the equations of motion which
are of the second order \cite{SVn}.

In this paper, we address the same issues for theories with higher unitary
gauge groups. In particular, we study the points 1 and 2 and show that the
results are basically the same: the real walls are always BPS saturated, and
 they
exist for all masses while the complex BPS equations have non-trivial solutions
only in the limited range of mass: $m \leq m_* = .28604\ldots \Lambda_{SQCD}$
for $SU(3)$, $m \leq m_* = .07539\ldots \Lambda_{SQCD}$
for $SU(4)$, etc. When $m > m_*$, the solutions are absent. A study of the 
corresponding equations of motion which would allow us to find the form and the
properties of non--BPS walls is now in progress.

The paper is organized as follows. In the next section we write down the
effective lagrangian for the model in interest, discuss its vacuum
structure and some generalities concerning the wall solutions. In Sect. 3,
we find the BPS solutions for the real walls. Complex walls in the Higgs
phase (in the limit $m \to 0$) are discussed in Sect. 4. In Sect. 5, we analyze
the case of small finite masses and find the solution as a series in a small
Born--Oppenheimer parameter. Numerical results for the complex walls for
 general values of mass are presented in Sect. 6. In Sect. 7, we discuss the
properties of a peculiar ``lower'' BPS solution at small masses.
Summary and some general 
discussion is the subject of the last section.

\section{The model}
\setcounter{equation}0

In this paper, we restrict ourselves with studying the model (\ref{LSQCD})
 with $N_f = N_c -1$. We also choose the simplest form for the mass matrix
$m_{ij} = m \delta_{ij}$. When $m$ is small (much smaller than 
$\Lambda_{\rm SQCD}$), the expectation values of the squark fields
$\langle s_{i} \rangle,\ \langle s'_{i} \rangle$ in  chirally asymmetric vacua 
are large. It turns
out that, for different flavours, the vacuum expectation values have different
colour orientations. Up to overall flavour and colour rotations, one may choose
 \cite{brmog}
  \beq
  \label{saver}
\langle s_{1} \rangle \ =\ \frac v{\sqrt{2}} \left(
\begin{array}{c} 1 \\0\\ \ldots\\0 \end{array} \right), \ldots ,\ \ 
  \langle s_{N_c -1} \rangle \ =\ \frac v{\sqrt{2}} \left(
\begin{array}{c} 0 \\ \ldots\\1\\0 \end{array} \right) \nonumber \\
\langle s'_{1}\rangle = \frac v{\sqrt{2}}(1,0,\ldots),\ \ldots, \ \ 
\langle s'_{N_c - 1} \rangle = \frac v{\sqrt{2}}(0,\ldots ,1,0)
   \eeq
with $v \propto m^{-1/2N_c}$; columns and rows display the colour structure.
 A set of Higgs averages (\ref{saver}) break the
 $SU(N_c)$ colour symmetry completely: all the gauge bosons acquire a large
mass. When $m$ is small, $v$ is large, the effective coupling $g(v)$ is 
 small and the
vacuum state has a trivial perturbative nature. One can say that the theory is
 in a weak--coupling Higgs phase.

The effective TVY lagrangian is written now in terms of the composite colorless
fields
 \beq
  \label{normN}
\Phi^3 = \frac 3{32\pi^2} {\rm Tr}\ W^2, \ \ {\cal M}_{ij} =\ 2 S'_iS_{j}
  \eeq
It is again a Wess--Zumino model with the superpotential
  \beq
{\cal W} = \frac{2}{3} \Phi^3 \left[ \ln \frac{\Phi^3 {\rm det} {\cal M}}
{\Lambda_{SQCD}^{2N_c + 1}} \ -\ 1 \right] - \frac{m}{2} {\rm Tr}\ {\cal M} 
\label{WTVY0}
\eeq  

 The superpotential is rigidly fixed from
the requirement that the conformal and the chiral anomaly of the 
theory (\ref{LSQCD}) under consideration are reproduced correctly.
For all vacuum states, the matrix $\langle s'_i s_j \rangle$ of the
vacuum expectation values is proportional to unity: 
$\langle s'_{i}s_j \rangle 
= C \delta_{ij}$. The same is true for the domain walls interpolating 
between different vacua. So, we can safely impose the constraint ${\cal M}_{ij}
 \ =\ X^2 \delta_{ij}$ and study the Wess--Zumino model with just two 
superfields $\Phi$ and $X$ and the superpotential
\beq
{\cal W} = \frac{2}{3} \Phi^3 \left[ \ln \{\Phi^3 X^{2(N-1)}\}
 \ -\ 1 \right] - \frac{m}{2} (N-1) X^2 
\label{WTVY}
\eeq
($N \equiv N_c$ and
we have set $\Lambda_{SQCD} \equiv 1$ ).
The corresponding potential for the lowest components $\phi$, $\chi$ of
the superfields $\Phi$, $X$ is
 \beq
U(\phi, \chi) \ =\ \left|\frac{\partial {\cal W}}{\partial \phi}
\right|^2 +
 \left|\frac{\partial {\cal W}}{\partial \chi}\right|^2 \ =\ 
4\left| \phi^2 \ln\{\phi^3 \chi^{2(N-1)}\} \right|^2 + (N-1)^2
\left|m\chi  - \frac{4\phi^3}{3\chi} \right|^2
\label{potTVY}
  \eeq

  The potential (\ref{potTVY}) has $N+1$ degenerate minima. One of them is chirally
symmetric $\phi = \chi = 0$ (one should take care, of course, that
the ratio $\phi^3/\chi$ would also tend to zero in the limit $\phi \to 0,\ \ 
\chi \to 0$). There are also $N$ chirally asymmetric vacua
   \beq
  \chi_k \ =\ \rho_*  e^{i\pi k/N},\ \ \ 
\phi_k \ =\ R_*  e^{-\frac{2i(N-1)\pi k}{3N}}\  
 \label{minn0}
  \eeq
with 
  \beq
\label{star}
\rho_* \ \equiv\ \ v\ =\   \left(\frac 4{3m} \right)^{1/(2N)},\ \ R_* 
\ =\ \left( \frac {3m}4 \right)^{(N-1)/(3N)}
  \eeq
The values of the superpotential (\ref{WTVY}) at the minima are
${\cal W}_{\rm sym} = 0$ and
  \beq
  \label{Wk}
{\cal W}_k \ =\ - \frac N2 \left( \frac {4m^{N-1}}3 \right)^{1/N}
e^{2\pi i k/N}
  \eeq
There are also the minima with the inverse sign of  $\chi$ (and the
appropriately chosen phase of $\phi$), but
they are physically the same as the minima (\ref{minn0}): the vacuum
expectation values of the gauge invariant operators $<{\rm Tr}\ \lambda^2>
\ = \ (32\pi^2/3) <\phi^3>$ and $<2 s'_i s_{j}> \ =\ <\chi^2> \delta_{ij}$
 are the same.  The vacuum values of the gluino condensate are 
$\langle {\rm Tr}\  \lambda^2 \rangle_{\rm sym} \ =\ 0$  and 
  
   \beq
 \label{condk}
\langle {\rm Tr}\ \lambda^2 \rangle_k \ =\ 8\pi^2  \left( \frac {4m^{N-1}}3 
\right)^{1/N} e^{2\pi i k/N}
  \eeq

To study the domain wall configurations, we should add to the potential
(\ref{potTVY}) the kinetic term which we choose in the simplest possible
form 
   \beq
\label{Lkin}
{\cal L}_{\rm kin} \ =\ |\partial \phi|^2 
+ |\partial \chi|^2
  \eeq 
(it follows from the term $1/4 \int d^4\theta \bar \Phi \Phi +
1/4 \int d^4\theta \bar X X$ in the superlagrangian)
 and solve the equations of motion with appropriate boundary conditions. 
 Following Refs.\cite{my,SV}, we will look in this paper only
 for the solutions of the more simple BPS equations (\ref{BPS}) which
are of the first order.

The model involves two types of walls. Some of them interpolate between the
chirally symmetric vacuum and a chirally asymmetric one. There are $N$ such
 walls which transform into each other inder trivial $Z_N$ phase rotation.
 It is
convenient to choose the chirally asymmetric vacuum with $k = 0$ in which case
the wall solution is purely {\it real}.

Other walls interpolate between chirally asymmetric vacua (\ref{minn0}) with
different $k$. We will restrict ourselves with studying the walls interpolating
between {\it adjacent} asymmetric vacua, say, between the vacuum with $k = 0$ 
and the vacuum with $k = 1$. For $N = 2,3$, it is the whole story but, starting
from $N = 4$, also the walls connecting the vacua with $k = 0$ and $k = 2$
etc. may appear. All such solutions are essentially {\it complex}.

The value of $\delta$ to be chosen in Eq.(\ref{BPS}) depends on the wall 
solution we are going to find. To fix it, note that the equations (\ref{BPS})
admit an integral of motion:
\beq
\label{ImW} 
{\rm Im} [ {\cal W}(\phi, \chi)e^{-i\delta}] \ =\ {\rm const}
\eeq
Indeed, we have
$$
e^{-i\delta} \partial_z {\cal W} = e^{-i\delta} \left(
\frac{\partial {\cal W}}{\partial \phi} \partial_z \phi  \ + \ 
\frac{\partial {\cal W}}{\partial \chi} \partial_z \chi \right) \ =  \\ 
\left| \frac{\partial {\cal W}}{\partial \phi} \right |^2 \ +\ 
\left| \frac{\partial {\cal W}}{\partial \chi} \right|^2 \ =\ 
e^{i\delta} \partial_z \bar {\cal W}
$$
  The real wall connects the vacua with ${\cal W} = 0$ and ${\cal W} = 
-N/2 (4m^{N-1}/3)^{1/N}$. These boundary conditions are consistent with 
Eq.(\ref{ImW}) only if $e^{i\delta} = \pm 1$ (the sign depends on whether the
walls goes from the symmetric vacuum to the asymmetric one when $z$ goes
from $-\infty$ to $+\infty$ or the other way round).

A complex wall connects the vacua where the superpotential (\ref{WTVY})
acquires non--zero values with different complex phases. For the wall
connecting the vacua with $k = 0$ and $k = 1$, the condition (\ref{ImW})
is consistent with the choice $\delta = \pi/N \pm \pi/2$ 
depending on the direction of the wall.

Bearing in mind Eqs. (\ref{eps},\ \ref{condk}), the energy densities of the BPS
walls are 
  \beq
 \label{epsr}
 \epsilon_r \ =\ N \left( \frac {4m^{N-1}}3  \right)^{1/N}
  \eeq
for the real walls and
  \beq
 \label{epsc}
 \epsilon_c \ =\ 2N \sin \frac\pi N \left( \frac {4m^{N-1}}3  \right)^{1/N}
\ =\ 2\epsilon_r \sin \frac \pi N
  \eeq
for the complex walls.

Before coming to grips with finding the numerical solutions of the BPS 
equations for an arbitrary value of mass, let us discuss what happens in the
limiting cases $m \to \infty$ and $m \to 0$ where the situation is 
considerably simplified. 

Consider first the case of large masses. In this case, one can integrate
the heavy matter fields out and write the effective lagrangian for the
composite chiral superfield $\Phi$. Technically, one should use the
 Born--Oppenheimer procedure and to freeze down the matter field $\chi$ so
 that the large second term in the potential (\ref{potTVY}) disappear.
 Proceeding in supersymmetric way, we get $X^2 = 4\Phi^3/3m$. Substituting
 it in the first term, we obtain the Veneziano--Yankielowicz effective
 lagrangian \cite{VY} which is the Wess--Zumino model for the single chiral 
 superfield $\Phi$ with the superpotential 
  \beq
  \label{WVY}
  {\cal W} \ =\ \frac{2N}3  \Phi^3 \left[ \ln  {\Phi^3}
  -\ 1 \right]
  \eeq 
where $\Phi$ is measured now in the units os
$$\Lambda_{SYM} = \left( \frac {3m}4 \right)^{(N-1)/3N}
 \Lambda_{SQCD}^{(2N+1)/3N}$$
  The corresponding  potential is
  \beq
  \label{potVY}
  U(\phi) =  |\partial {\cal W}/\partial \phi|^2 = \ 4N^2 |\phi^4|
|\ln \phi^3|^2
  \eeq
This expression 
 is not yet well defined: the logarithm has many sheets, and one should specify
 first what particular sheet should be taken. An accurate
 analysis taking into account the fact that the topological charge 
  $\nu \ \propto\
 \int {\rm Tr} \{G^2 \tilde G^2 \} d^4x $ 
 in the original theory
(\ref{LSQCD}) is quantized to be integer reveals that the true potential
is glued out of $N$  such sheets \cite{Kovner,my}. 
\footnote{ Quite an analogous situation
  holds in the Schwinger model: the true bosonized lagrangian (in
   the Schwinger model it is just {\it equivalent} to the original theory )
   is glued out of several branches when taking into account the effects
   due to quantization of topological charge \cite{QCD2}.}
The gluing occurs when the phase
of the expression under logarithm reaches the values $\pm \pi/N$
 \cite{Kovner,my}:
  \beq
  \label{glue}
\ln \phi^3 & \equiv & \ln |\phi^3| + i{\rm arg} (\phi^3), \ \ \ \ 
{\rm arg} (\phi^3) \ \in \ \left(- \frac \pi N, \frac \pi N \right) 
\nonumber \\
\ln \phi^3 & \equiv & \ln |\phi^3| + i\left[{\rm arg} (\phi^3) - 
\frac {2\pi} N \right],
 \ \ \ \ 
{\rm arg} (\phi^3) \ \in \ \left(\frac \pi N, \frac {3\pi} N \right) 
\nonumber \\
&  & \ldots \nonumber \\
\ln \phi^3 & \equiv &\ln |\phi^3| + i \left[{\rm arg} (\phi^3) - 
\frac{2(N-1)\pi}N \right], \nonumber \\ 
& & {\rm arg} (\phi^3)  \in  \left(\frac{\pi(2N-3)}N, \frac{\pi(2N-1)}N 
\right) 
  \eeq
 (remind that only the field $\phi^3 \propto {\rm Tr}\ \lambda^2$, not
  $\phi$ itself has a direct physical meaning).
Bearing the prescription (\ref{glue}) in mind, the potential (\ref{potVY})
has $N+1$ minima as it should: $\phi_{\rm sym} = 0$ and 
$\phi_k = e^{2i\pi k/3N}$. 

As was shown in Ref.\cite{my}, the BPS equations involve only the  real
solutions with the chirally symmetric vacuum at one of the boundaries
 in this case. Non-trivial complex domain walls connecting
different chirally asymmetric vacua are absent.

Consider now the case of small masses.
In this case, chirally
asymmetric vacua are characterized by large expectation values of the matter 
scalar field $\langle \chi \rangle \sim \rho_* \propto m^{-1/2N}$ . 
Again,  the theory involves two different energy scales, and one
can tentatively integrate out heavy fields  and to write down the Wilsonean
effective lagrangian describing only light degrees of freedom. Proceeding
in the Born--Oppenheimer spirit, we should freeze now the heavy field 
$\phi$ in the potential (\ref{potTVY}) so that the large first term in the
potential acquire its minimum (zero) value. In contrast to the
large mass situation, this can now be achieved in two ways: either
by setting $\phi = 0$ or by setting $\phi^3 \chi^{2(N-1)} = 1$. In the first
 case,
we will obtain the effective lagrangian describing the dynamics of the 
chirally symmetric phase which is just the lagrangian of free light chiral
field $X$ involving $\chi$  and its superpartner.

 The second choice results
in the lagrangian describing the dynamics of the chirally asymmetric phases.
It is
the lagrangian of the Wess--Zumino model with a single chiral
superfield $X$ and a non-trivial superpotential
  \beq
{\cal W} = -\frac{2}{3X^{2(N-1)}} -\frac{m}{2} (N-1) X^2
\, .
\label{WHiggs}
\eeq
This lagrangian is well known and was obtained earlier from  
instanton and/or from holomorphy
considerations \cite{brmog}.
The corresponding potential $U \ =\ |\partial W/\partial \chi |^2$ has
$N$ different non--trivial minima at
 $ \chi_k \ =\  \rho_* e^{i\pi k/N}$. 
When $m \ll 1$, a large expectation value $\langle \chi \rangle $ results
 in breaking down 
the gauge symmetry of the original theory by the Higgs mechanism: 
the theory is in the Higgs phase.

The BPS equations corresponding to the superpotential (\ref{WHiggs})
admit non-trivial complex domain wall solutions connecting different asymmetric
vacua. When $N=2$, the solution can be found analytically \cite{my}. For
$N \geq 3$, there are only numerical solutions which will be presented and
discussed in Sect. 4 of the paper.

  In this approach, we are not able, however,  to study the real domain walls
interpolating between the
chirally symmetric and a chirally asymmetric vacua. Such a wall corresponds
to going through a high energy barrier separating two kind of vacua. It is
a remarkable consequence of supersymmetry and of the related BPS condition
(\ref{eps}) that the energy of such a wall is still not large. For $N=2$,
such walls were analyzed in Ref.\cite{my}. In the next section, we do the same
for arbitrary $N$ and reach the same conclusions.

Thereby, the physical situation in the limits $m \to 0$ and $m \to \infty$
is somewhat different. In both cases, we have $N+1$ different vacua. However,
an analog of nontrivial complex walls  connecting different
asymmetric vacua which are present at small masses, is absent when the mass
is large. In the latter case, only real walls are present. It is therefore
very interesting to understand what happens in between, at intermediate
 values
of masses, and how the transition from one regime to another  occurs.
That was the main motivation for our study.
   
 All the calculations were performed in the framework of the effective theory
with the  potential (\ref{potTVY}). The status of this effective 
theory is somewhat more
uncertain than that of (\ref{WHiggs}) --- for general value of mass,
the TVY effective lagrangian is not Wilsonean; light and heavy degrees
of freedom are not nicely separated. But it possesses all the relevant
symmetries of the original theory 
 and satisfies the anomalous Ward identities for correlators at zero
momenta. We think that the use of the TVY lagrangian is justified as
far as the vacuum structure of the theory is concerned.
  
\section{Real Walls}
\setcounter{equation}0
The BPS equations (\ref{BPS}) with the superpotential (\ref{WTVY}) have the 
form
  \beq
\label{fihi}
 \phi '\  = \ e^{i\delta} \cdot 2\bar \phi^2 \ln\{\bar\phi^3
\bar \chi^{2(N-1)}
 \} \nonumber \\
 \chi ' \ =\ e^{i\delta} \cdot (N-1) \left[
 \frac{4\bar\phi^3}{3\bar\chi} - m\bar\chi \right]
  \eeq
($O' \equiv \partial_z O$).
To find the wall interpolating between $\phi = \chi = 0$ at $z = -\infty$
and $\phi = R_*, \chi = \rho_*$ at $z = \infty$, we have to choose
$\delta = \pi$ (or $\delta = 0$ for the wall going in the opposite direction).
With this choice and the boundary conditions given, the solutions $\phi(z)$
and $\chi(z)$  are going to be real so that we have a simple system of just
two first--order differential equations.

For all $N$, the dynamics of this system is quite similar to that
in the case $N=2$ studied in Ref.\cite{my}. The solution exists for all
masses. For large $m$, the heavy matter field can be integrated out, and we
arrive at the BPS equation for the supersymmetric gluodynamics
  \beq
 \label{fiN}
\phi' \ =\ -2N \phi^2 \ln \left(\phi^3/R_*^3\right)
  \eeq
The solution of this equation with the boundary conditions
 $\phi(-\infty) = 0,\ \phi(\infty) = R_*$ can be expressed into integral
logarithms.

As was the case for $N=2$, the solution of the system (\ref{fihi}) can be
found in an analytic form also for small masses. The wall trajectory consists
then in two distinct regions. On the first stage, only the light field $\chi$
is changed, $\chi(z) = \rho_* e^{mz}$, while the field $\phi$ stays frozen at 
zero. Then, at $z=0$, the trajectory abruptly turns. $\chi(z)$ stays frozen
at its vacuum value $\rho_*$ and the equation for $\phi(z)$ is reduced to
$\phi' \ =\ -2 \phi^2 \ln \{\phi^3 \rho_*^{2(N-1)}\}$ which is the same as
Eq.(\ref{fiN}). The second stretch is much thinner than the first one, its
width being of order $R_*^{-1} \propto m^{-(N-1)/3N}$. It is on this second
stage when the high and narrow potential barrier between the regions 
$\phi \sim 0$ and $\phi \sim \chi^{-2(N-1)/3}$ is penetrated. The width of the
first stretch is $\sim 1/m$ and it carries the fraction $(N-1)/N$ of the total
energy (\ref{epsr}). Correspodningly, the second thin stretch carries the
fraction $1/N$ of the total energy.

\begin{figure}
  \begin{center}
        \epsfxsize=300pt
        \epsfysize=400pt
        \vspace{-5mm}
        \parbox{\epsfxsize}{\epsffile{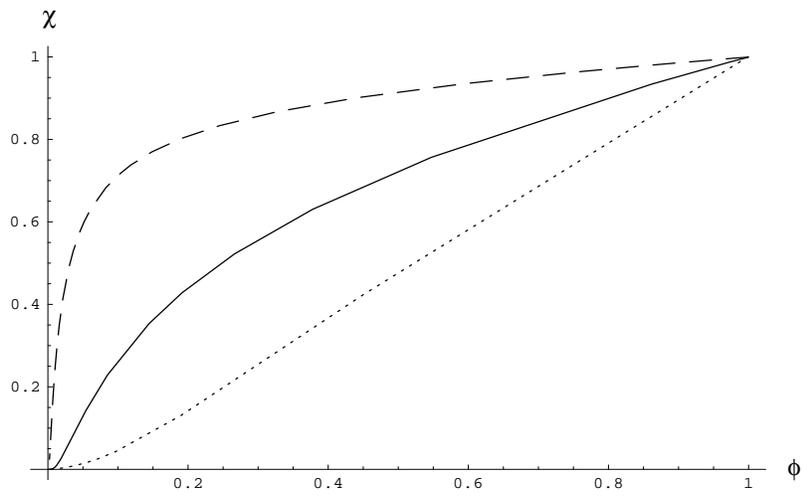}}
        \vspace{5mm}
    \end{center}
\caption{Real walls ($N=3$). The parametric plots for $m=1$ (solid line),
$m = .1$ (dashed line), and $m = 10$ (dotted line). $\phi$ and $\chi$ are
measured in units of $R_*$ and $\rho_*$, respectively. }
\label{reals}
\end{figure}

In the intermediate range of masses, the solution has to be found numerically.
Parametric plots in the ($\phi, \ \chi$) plane for $N = 3$ and 
 different values of $m$ are drawn in Fig. \ref{reals}.

 \section{Domain Walls in Higgs Phase}
\begin{figure}
  \begin{center}
        \epsfxsize=300pt
        \epsfysize=400pt
        \vspace{-5mm}
        \parbox{\epsfxsize}{\epsffile{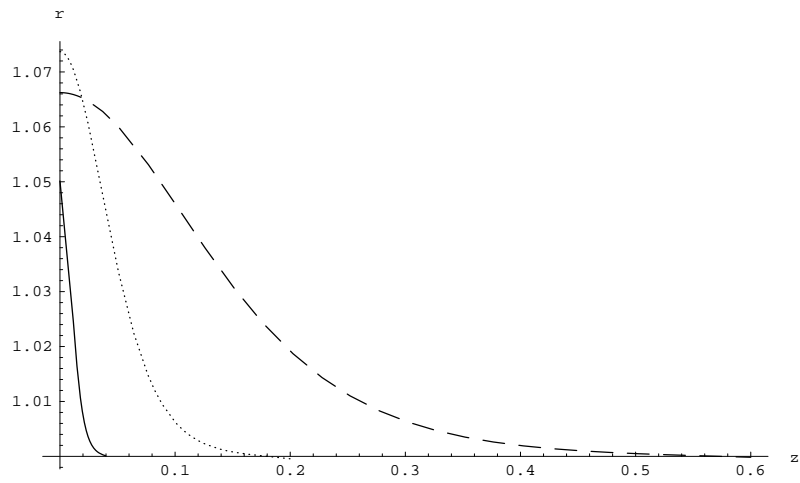}}
        \vspace{5mm}
    \end{center}
\caption{BPS walls in Higgs phase for $N = 3$ (dashed line), $N = 5$
(dotted line) and $N = 10$ (solid line). }
\label{prof0s}
\end{figure}

 \setcounter{equation}0
We have got to solve the BPS equations
  \beq
  \label{BPSHig}
 \chi ' \ =\ e^{-i \pi(N-2)/2N} 
\frac{\partial \bar {\cal W}}{\partial \bar \chi}\, ,
    \eeq
where ${\cal W}$ is the superpotential (\ref{WHiggs}), with the boundary
conditions
 \beq
 \label{bcHgs}
\chi(-\infty) \ =\ \rho_*;
\ \ \ \ \ \  \chi(\infty) \ =\  \rho_*e^{i\pi/N}\, ;
  \eeq
  It is convenient to introduce polar variables $\chi \ =\ \rho e^{i\alpha}$. 
The equations (\ref{BPSHig}) acquire the form
  \beq
\label{rogam}
 \rho ' \ =\ (N-1) \left\{ m\rho \sin \gamma \ -\ \frac 4{3\rho^{2N-1}}
\sin [\gamma(N-1)] \right\} \nonumber \\
 \gamma '\ =\ 2(N-1) \left\{ m \cos \gamma \ + \ \frac 4{3\rho^{2N}}
\cos [\gamma(N-1)] \right\}
  \eeq
 where $\gamma \equiv 2\alpha - \pi/N$ changes from $\gamma = -\pi/N$ at
$z = -\infty$ to $\gamma = \pi/N$ at $z = \infty$.
 For $N = 2$, these equations were solved in Ref.\cite{my}. The solution is
analytic:
  \beq
  \label{rogamsol}
\rho(z) \ &=&\ \rho_* \nonumber \\
\tan \gamma(z) \ &=&\ \sinh [4m(z-z_0)]
  \eeq
or in the complex form:
  \beq
 \label{chisol}
\chi(z) \ =\ \rho_* \frac {1 + i e^{4m(z - z_0)}}{\sqrt{1 + e^{8m(z-z_0)}}}
  \eeq
($z_0$ is the position of the wall center).
For $N \geq 3$, the solutions can be found numerically. The profiles for the
ratio
$r(z) = \rho(z)/\rho_*$ in the interval $z_0 \equiv 0 \leq z < \infty$
[it is a half of the wall, another half being restored by symmetry
considerations: $\rho(-z) = \rho(z)$]
with different $N = 3, 5 , 10$ are presented in Fig. \ref{prof0s}.

We see that the dependence $\rho(z)$ is not flat anymore but displays a bump
in the middle. To understand it, remind that the system (\ref{rogam})
has the integral of motion (\ref{ImW}). In our case, it amounts to
  \beq
 \label{intmot}
\frac{m(N-1)}2 \rho^2 \cos \gamma \ -\ 
\frac2{3 \rho^{2(N-1)}} \cos [ \gamma (N-1)]  \ =\  
\frac N2 \left( \frac{4m^{N-1}}3 \right)^{1/N} \cos \frac \pi N 
   \eeq
as follows from the boundary conditions (\ref{bcHgs}) and the relations
(\ref{Wk}). In the middle of the wall, $\gamma = 0$, and  the condition
(\ref{intmot}) implies
 \beq
  \label{xN}
(N-1) x^2 - \frac 1{x^{2(N-1)}} \ =\ N \cos \frac \pi N
  \eeq
( $x \equiv r(0)$). It is not difficult to
observe that the real root of the algebraic equation (\ref{xN}) is slightly
greater than 1 for $N \geq 3$. When $N$ is large, $x-1$ tends to zero 
$\propto 1/N$.

\section{Born -- Oppenheimer Expansion at Small Masses}
\setcounter{equation}0

When $m \neq 0$, the gauge degrees of freedom associated with the superfield
$\Phi$ do not decouple completely and should be taken into account. The full
system (\ref{fihi}) of the BPS equations for the complex domain walls has the
form
   \beq
   \label{4sys}
    \begin{array}{l}
 \rho '\ =\ (N-1) \left[ m\rho \sin (2\alpha - \pi/N) -
 \frac{4R^3}{3\rho} \sin (3\beta - \pi/N) \right]
\\
 \alpha '\ =\ (N-1) \left[ m \cos (2\alpha - \pi/N) -  
\frac{4R^3}{3\rho^2} \cos (3\beta - \pi/N) \right]
 \\
 R '\ =\ -2R^2 \left[ \sin (3\beta - \pi/N) \ln(R^3\rho^{2(N-1)}) +
\cos (3\beta - \pi/N) [3\beta + 2\alpha (N-1)] \right]  \\
 \beta '\ =\ 2R \left[- \cos(3\beta - \pi/N) \ln(R^3\rho^{2(N-1)}) +
\sin (3\beta - \pi/N) [3\beta + 2\alpha (N-1)] \right] 
\end{array} 
   \eeq
 where, as in the previous section, we have chosen $\delta \ =\ \pi/N -\pi/2$
and introduced the polar variables $\chi = \rho e^{i\alpha},\ 
\phi = R e^{i\beta}$. One should solve the system (\ref{4sys}) with the 
 boundary conditions
  \beq
  \label{4bc}
\rho(-\infty)  =\rho(\infty) = \rho_*; \ \ R(-\infty) = R(\infty) = R_*; 
\nonumber \\
\alpha(-\infty) = \beta(-\infty) = 0;\ \ \alpha(\infty) = \pi/N;\ \ 
\beta(\infty) = - \frac{2(N-1) \pi}{3N}
  \eeq
 When $N=2$, the system (\ref{4sys}) is reduced to that studied in 
Refs.\cite{SV,SVn}.

 Generally, we should solve the full system (\ref{4sys})
numerically. The problem is simplified when noting that the wall 
 solution should be symmetric with respect to its center. 
Let us seek for the solution centered at $z=0$ so that 
\beq
\label{sym}
\rho(z)  = \rho(-z), \ R(z) = R(-z),\ \nonumber \\
\alpha(z)  = \pi/N - \alpha(-z),
\ \beta(z)  = -2(N-1)\pi/(3N) - \beta(-z)
\eeq
Indeed, one can be easily convinced that the Ansatz (\ref{sym}) goes through
the equations (\ref{4sys}). 
It is convenient to solve the equations (\ref{4sys}) numerically on the
half--interval from $z=0$ to $z = \infty$. The symmetry (\ref{sym}) 
dictates $\alpha(0) = \pi/(2N),\ \beta(0) = -(N-1)\pi/(3N)$. 
The values $\rho(0)$ and $R(0)$ are related by the condition (\ref{ImW})
which reads here
  \beq
  \label{Rrho}
\frac{4R^3(0)}3 \left\{ \ln[R^3(0)\rho^{2(N-1)}(0)] - 1\right \} + m\rho^2(0) 
(N-1)\ =\ N \left( \frac {4m^{N-1}}3 \right)^{1/N} \cos \frac \pi N 
   \eeq
Thus, only one parameter at $z=0$ [say, $R(0)$] is left free. We should fit
it so that the solution would approach the values specified in Eq.(\ref{4bc})
at $z \to \infty$.
 The numerical solution for this problem will be presented and discussed
 in the next section. Here we will concentrate on the case when $m$ is
small where some analytic results can be obtained.

In the limit $m \to 0$, we can just freeze the heavy variables: 
   \beq
 \label{freeze}
\beta = 
-2\alpha(N-1)/3;\ R = \rho^{-2(N-1)/3}
  \eeq
 in which case the first two equations 
in Eq.(\ref{4sys}) reproduce the system (\ref{rogam}) studied in the previous 
section.  This is the leading order of the Born--Oppenheimer expansion.
To proceed further, we should allow for the fast variables $R,\ \beta$
to deviate from their zero order values (\ref{freeze}). Integrating these
deviations out, one can obtain the corrections to the leading order
Born--Oppenheimer hamiltonian for the slow variables $\rho,\ \alpha$.
The spectrum of the low--energy effective hamiltonian and all other
quantities of interest can be expanded in a series  over a small parameter.
In our case, the relevant expansion parameter is $\sim m/R_* \propto
m^{(2N+1)/3N}$.

Such an expansion can be carried out also on the level of dynamic equations.
One should present
$$ \rho \ =\ \rho_0 + \rho_1 + \rho_2 + \ldots,\ \ \ 
  \alpha \ =\ \alpha_0 + \alpha_1 + \alpha_2 + \ldots,\  \\
R \ = \ R_0 + R_1 + R_2 + \ldots, \ \ \ \beta \ =\ \beta_0 + \beta_1
+ \beta_2 + \ldots $$
 , where $\rho_0$ and $\alpha_0$ are the solutions of the system
(\ref{rogam}) and $R_0,\ \beta_0$ are found from Eq.(\ref{freeze}), 
and linearize the equations in each subsequent order. Let us do it for the
BPS system (\ref{4sys}). We will be particularly interested in the quantity
$R(0)$ to compare it with the numerical results of the next section. 

Let us look first at the two last equations in Eq.(\ref{4sys}). In the first
non-trivial order, we have
  \beq
 \label{linRb}
\begin{array}{l}
R_0' = -2R_0^2 \left\{ \sin(3\beta_0 - \pi/N) \left[ \frac {3R_1}{R_0} 
+ \frac{2(N-1)\rho_1}{\rho_0} \right] + \cos(3\beta_0 - \pi/N) [3\beta_1
+ 2\alpha_1(N-1)] \right\} \\
\beta_0' = 2R_0 \left\{ - \cos(3\beta_0 - \pi/N) \left[ \frac {3R_1}{R_0} 
+ \frac{2(N-1)\rho_1}{\rho_0} \right] + \sin(3\beta_0 - \pi/N) [3\beta_1
+ 2\alpha_1(N-1)] \right\} \\
\end{array}
\eeq
  From this, one can readily express $R_1$ and $\beta_1$ to be substituted in
the linearized version of the first pair of the equations in Eq.(\ref{4sys}).
As a result, we obtain the linear system for $\rho_1$ and  
$\gamma_1 = 2\alpha_1 - \pi/N$ :
 \beq
 \label{lin1}
\rho'_1 \ &=&\ (N-1) \left\{ - \frac{4(N-1)}9 \rho_0^{-2(2N+1)/3} \rho_0'
+ \rho_1\left[ m \sin \gamma_0 + \frac{4(2N-1)}{3\rho_0^{2N}} 
\sin[\gamma_0(N-1)] \right] + \right. \nonumber \\
& &\left. \gamma_1 \left[ m\rho_0 \cos \gamma_0 - \frac{4(N-1)}{3\rho_0^{(2N-1)}}
\cos[\gamma_0(N-1)] \right] \right\} \nonumber \\
  \gamma'_1 \ &=&\ 2(N-1) \left\{ - \frac{2(N-1)}9 \rho_0^{-2(2N+1)/3} \gamma_0'
 - \frac{8N}{3\rho_0^{(2N+1)}} \cos[\gamma_0(N-1)] \rho_1 - \right. \nonumber \\
& &\left. - \gamma_1\left[ m \sin \gamma_0 + \frac{4(N-1)}{3\rho_0^{2N}} 
\sin[\gamma_0(N-1)] \right] \right\}\, , \nonumber \\
  \eeq
 The proper initial conditions for this
system at the middle of the wall are
  \beq
 \label{ini1}
\rho_1(0) = 0,\ \ \ \ \gamma_1(0) = 0
  \eeq
The fact that the value of the phase in the middle is not shifted is a trivial
corollary of the symmetry conditions (\ref{sym}). To understand why also
$\rho(0)$ is not shifted, one should look at the relation (\ref{Rrho}). 
Linearizing it with respect to $R(0)$, $\rho(0)$, we observe that the partial
derivative of the left--hand side over $R(0)$ is zero for the 
$0^{\underline{th}}$ order solution while the derivative over $\rho(0)$ is
not. Hence, $\rho_1(0) = 0$, indeed. When the initial conditions (\ref{ini1}) are
posed, the solutions $\rho_1(z),\ \alpha_1(z)$ of the system
(\ref{lin1}) should approach zero at $z \to \infty$. We checked numerically 
that they do.

Bearing in mind that $\rho_1(0) = 0$, it is easy to find what $R_1(0)$ is. 
For $z = 0$, the second equation in (\ref{linRb}) is reduced to
  \beq
  \label{R10}
R_1(0) \ =\ \frac{\beta_0'(0)}6 \ =\ - \frac{(N-1)^2}9 
\left[ m + \frac 4{3\rho_0^{2N}(0)} \right]
 \eeq
where the relation (\ref{freeze}) and the second equation in Eq.(\ref{rogam})
were used. To find the shift in the next order, one should expand the 
equation for $\beta'(0)$ further. We obtain
  \beq
 \label{R2prom}
\beta'_1(0)\ =\ 6R_2(0) \ +\ \frac{3R_1^2(0)}{R_0(0)}\  + 
\frac {4(N-1) R_0(0)}{\rho_0(0)} \rho_2(0)
 \eeq
The second order shift in $\rho(0)$ can be found from the expansion of 
Eq.(\ref{Rrho}):
  \beq
 \label{rho2}
 \rho_2(0) \ =\ - \frac{3R_0(0) R_1^2(0)}{(N-1) \rho_0(0)
[m +  \frac{4R_0^3(0)}{3\rho^2_0(0)}]}
  \eeq
 $\beta'_1(0)$ can be found by   differentiating the solution of the 
linear algebraic system (\ref{linRb}) at $z = 0$. Using Eqs.(\ref{rogam}, 
\ref{freeze}) again and combining everything, we arrive at the final result
  \beq
 \label{etam}
\eta(m) \ =\ \frac{R(0)}{R_0(0)} \ =\ 1\  - \ \frac{(N-1)^2}{9R_0(0)} 
\left[ m + \frac 4{3\rho_0^{2N}(0)} \right] - \nonumber \\
\frac{(N-1)^3}{162 R_0^2(0)} 
\left[ m + \frac 4{3\rho_0^{2N}(0)} \right]
\left[ m(7N-1) - \frac{32(N-1)}{3\rho_0^{2N}(0)} \right] \ + \nonumber \\
O[m^3/R_0^3(0)]
  \eeq
For $N=2$ when $\rho_0(0) = \rho_*$ and $R_0(0) = R_*$, the expression
simplifies \cite{SVn}:
 \beq
  \label{etam2}
  \eta(m) \  =\ 1 - \frac 29 \left(\frac {4m^5}3 
\right)^{1/6} - \frac 5{81}  \left(\frac {4m^5}3 \right)^{1/3} + O(m^{5/2})
  \eeq

\section{Two BPS solutions and Phase Transition in Mass}
\setcounter{equation}0
  Whem $m$ is neither too large nor too small, the Born--Oppenheimer
approximation does not apply and we are in a position to solve the full 
system of 4 equations (\ref{4sys}) numerically. We did it for $N = 3$ and
$N = 4$. As was mentioned in the previous section, we use the symmetry 
relations (\ref{sym}) which fix the initial conditions at the middle of the
wall for the phases and the relation (\ref{Rrho}) between $\rho(0)$ and $R(0)$
so that only one free parameter is left. We fit it so that the solution
would tend to the vacuum values (\ref{4bc}) when $z \to \infty$.

\begin{figure}
\grpicture{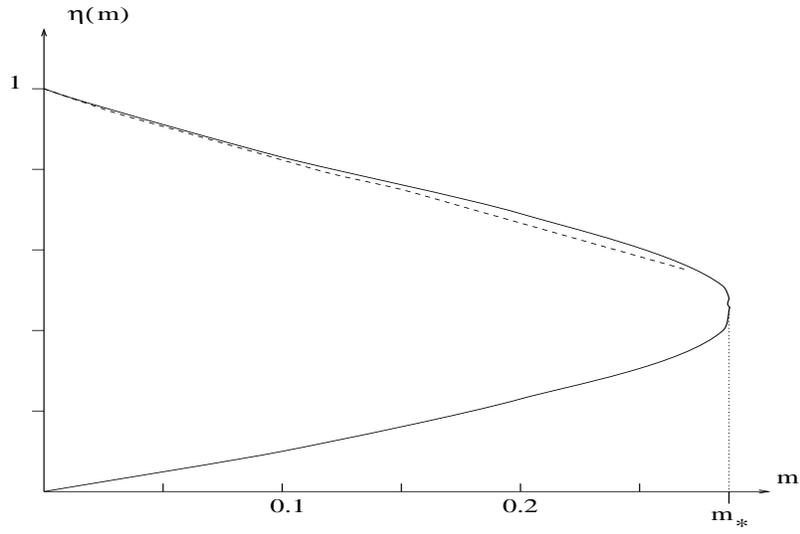}
\caption{The 
ratio $\eta = R(0)/R_0(0)$  as a function of mass for the 
$SU(3)$ theory. The dashed
line describes the analytic result (\ref{etam}) valid for small masses.}
\label{plot3s}
\end{figure}

\begin{figure}
\grpicture{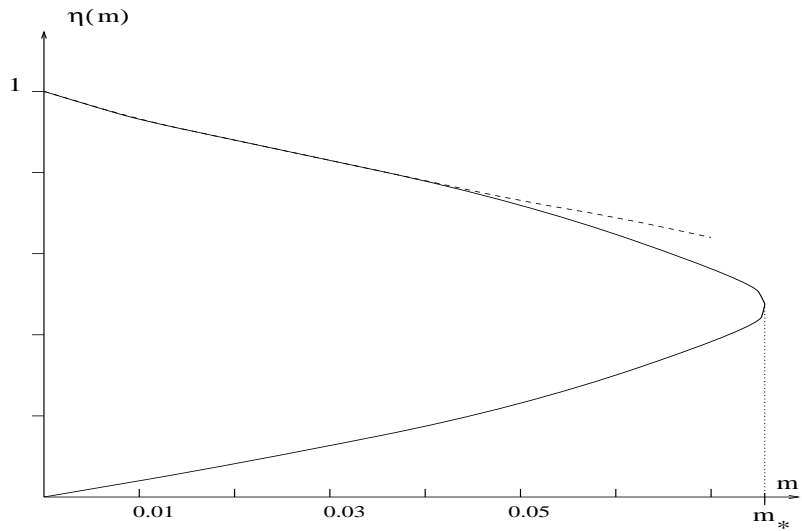}
\caption{The same as Fig. \ref{plot3s} for the $SU(4)$
theory.}
\label{plot4s}
\end{figure}

 The situation turned out to be pretty much analogous to what happens at
$N = 2$. First of all, the solutions exist only in a limited
range of masses. When $m$ is larger than some critical value $m_*$, the
integral trajectory always {\it misses} the vacuum (\ref{4bc}) no matter
what value for $R(0)$ is chosen. Second, when $m < m_*$, there are not one, but
{\it two} different solutions with a larger and a smaller value of $R(0)$.
In Figs. \ref{plot3s}, \ref{plot4s} we plotted the dependence of 
$\eta = R(0)/R(\infty)$ on $m$  for both branches. 
 We see that, at $m = m_*$, two branches are joined together. This {\it is}
 the reason why no solution exists at larger masses.
\footnote{The presence of two solutions at $m < m_*$ and their absence
at $m > m_*$ can be naturally understood by making some simple observations
on the phase portrait of the system (\ref{4sys}) \cite{SVn}.}

   The value of the critical mass $m_*$ for higher
unitary groups turned out to be much smaller than that for the $SU(2)$ theory:
 \beq
\label{mcrit}
m_*^{SU(3)} \ =\ .28604\ldots;\ \ \ \ \ \ 
m_*^{SU(4)} \ =\ .07539\ldots
 \eeq
as compared to $m_* = 4.67059\ldots$ for $SU(2)$.

\begin{figure}
  \begin{center}
        \epsfxsize=300pt
        \epsfysize=400pt
        \vspace{-5mm}
        \parbox{\epsfxsize}{\epsffile{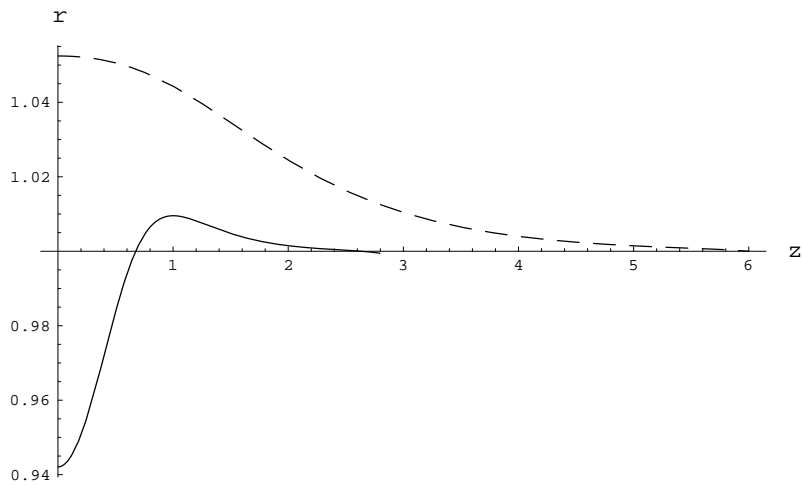}}
        \vspace{5mm}
    \end{center}
\caption{ The ratio $r(z) =  \rho(z)/\rho_*$ in the
$SU(3)$ theory as a function of $z$
 for $m = m_*$ (solid line) and for  the upper BPS branch at $m = .1$ 
(dashed line). }
\label{rho3s}
\end{figure}

 Let us discuss
 now what happens with these two branches in the small mass limit. Consider
 first the upper branch. 
 The illustrative profiles  $\rho(z)/\rho_*$ in the $SU(3)$ theory for
the upper branch at $m = .1$ and for
$m = m_* = .28604$ (when the solution is unique)  are plotted in Fig.
 \ref{rho3s}.  
Comparing the dashed curves in Fig. \ref{rho3s} and Fig. \ref{prof0s}, we see
 that, for small masses,
  the solution approaches, as it should, the ``Higgs wall''  solution 
studied in Sect. 4. 
 The numerical findings for $\eta(m)$ are in an excellent agreement with
the approximal analytic result (\ref{etam}) both for $SU(3)$ and for $SU(4)$. 

  The result (\ref{etam}) allows one to understand why the values of $m_*$
fall down with $N$ and to make an estimate of $m_*$ at large $N$. Indeed,
the critical mass $m_*$ and the value of the mass where the Born--Oppenheimer
expansion in Eq.(\ref{etam}) breaks down should be of the same order.
For large $N$, the expansion parameter is $\kappa \sim N^2(m/R_*) \sim
N^2 m^{2/3}$. Assuming $\kappa \sim 1$, we arrive at the conclusion that
$m_*$ falls down as $$m_*(N) \ \propto\ N^{-3}$$ in the limit $N \to \infty$.

  The lower BPS branch is something new which was not and could not be seen
in the framework of the effective Higgs theory (\ref{WHiggs}). The latter
was obtained by freezing down to zero the logarithm in the full potential
(\ref{potTVY}). But, as was already mentioned, we could equally well set
$\phi  = 0$. Actually, we see, indeed, that $R(0) \to 0$ when $m \to 0$
for the lower branch, i.e. the solution passes in the vicinity of the
chirally symmetric vacuum on its way from one chirally asymmetric vacuum
to another.

A similar lower BPS solution exist also for $N = 2$, but in that case it is 
much simpler approaching in the limit $m \to 0$
 a combination of two {\it real} walls separated
at a large distance (The phases $\alpha(z)$ and $\beta(z)$ are changed, of
course, so that the whole solution is complex, but the change occurs in the 
central region where the absolute values of the fields $R$ and $\rho$ are
very small.).
For $N \geq 3$, it cannot be true because the energy of a complex BPS wall is 
not just twice the energy of a complex BPS wall, but is lower. One can say
that, at small masses,
 two real BPS walls with the energy (\ref{epsr}) attract each other at large
 distances and two different
bound states with the same energy (\ref{epsc})
  are formed.
\footnote{This terminology looks somewhat artificial in 4 dimensions where
the walls have infinite energy involving the area factor, but one could 
equaly well discuss a 
2--dimensional ${\cal N} = 2$ supersymmetric Wess--Zumino model with the 
superpotential
(\ref{WTVY}) in which case the walls are interpreted as solitons and
the wall spectrum is just {\it the} spectrum, at least an essential part of
it (there may be also some breathers). E.g., for $N=3$ and $m < m_*$, we have 
in our disposal 6 real bosonic
solitons and antisolitons with the mass $\epsilon_r$ and twice as much 
 complex bosonic solitons and antisolitons with the mass 
$\sqrt{3} \epsilon_r$, each soliton possessing a 
superpartner associated with the fermion zero mode dwelling on the wall.
   One may wonder whether
the model is exactly solvable in Zamolodchikov's sense ? }

\section{Lower BPS branch at small masses}
\begin{figure}
\begin{center}
        \epsfxsize=300pt
        \epsfysize=400pt
        \vspace{-5mm}
        \parbox{\epsfxsize}{\epsffile{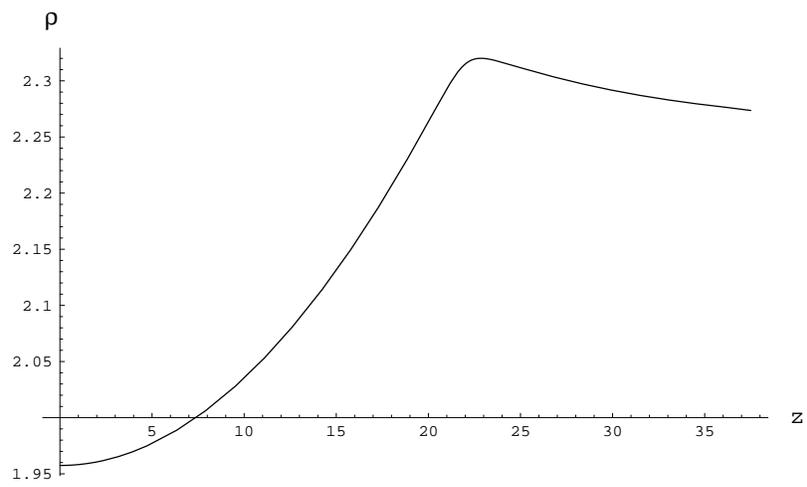}}
        \vspace{5mm}
    \end{center}
\caption{Profile $\rho(z)$ of the lower BPS branch at $N=3,\ m=0.01$}
\label{lrho3s}
\end{figure}

\begin{figure}
\begin{center}
        \epsfxsize=300pt
        \epsfysize=400pt
        \vspace{-5mm}
        \parbox{\epsfxsize}{\epsffile{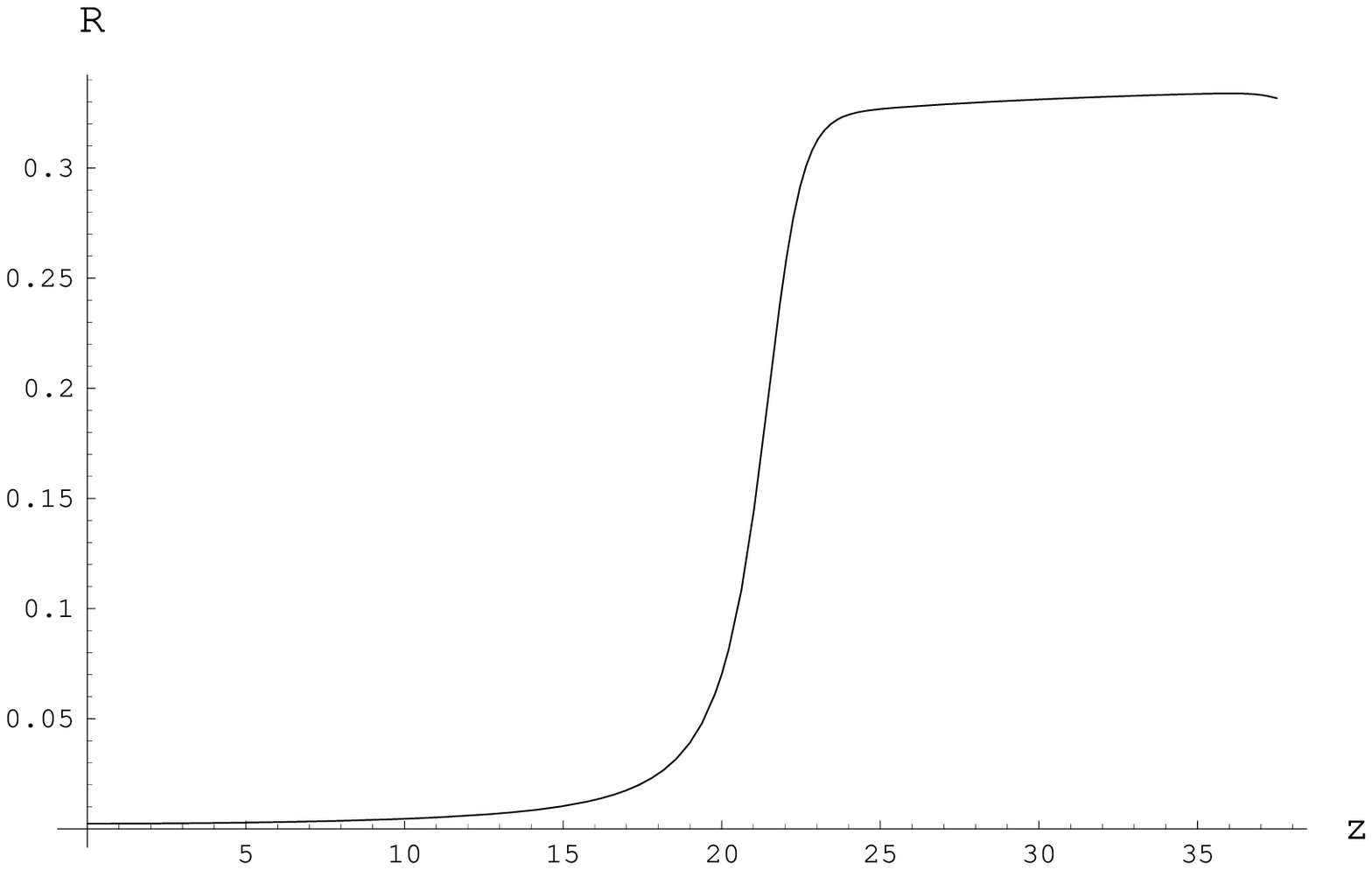}}
        \vspace{5mm}
    \end{center}
\caption{Profile $R(z)$ of the lower BPS branch at $N= 3,\ m=0.01$.}
\label{lRs}
\end{figure}

\setcounter{equation}0
  The properties of this new complex BPS branch are rather peculiar. They
are worth studying in some details. 
As was also the case for the real walls, this solution penetrates a high
and narrow potential barrier so that the central region of the wall finds
itself in the vicinity of the chirally symmetric vacuum. However, in
contrast to the case $N = 2$, it always passes the minimum $\phi = \chi 
= 0$ at a finite distance: though $\lim_{m \to 0} R(0)$ is zero, 
 $\lim_{m \to 0} \rho(0)$ is not. This can be immediately seen from the 
relation (\ref{Rrho}). For $N \geq 3$, the right--hand side is non--zero, and,
when $R(0) \to 0$, $\rho(0)$ tends to the value
  \beq
 \label{rho0}
\rho_{m = 0}(0) \ \equiv \ \rho_0\ =\ \sqrt{\frac{N \cos \frac \pi N}{N-1}} 
\left( \frac 4{3m} \right)^{1/2N} \ =\ \sqrt{\frac{N \cos \frac \pi N}{N-1}}
\rho_*
   \eeq
Let us look first at the numerical solution for $\rho(z)$ and $R(z)$ for the
lower BPS wall at $N = 3,\ m = .01$ shown in Figs. \ref{lrho3s}, \ref{lRs}.
 We see that
the wall consists of three distinct regions. First, when going out from the 
middle point, $\rho(z)$ goes gradually up while $R(z)$ stays practically
at zero. Then, in the intermediate region, $R(z)$ changes rather abruptly
while $\rho(z)$ is practically not changed. Finally, $\rho(z)$ and $R(z)$
change together in a smooth concerted way until the trajectory levels off
at the asymmetric vacuum values $\rho_*$ and $R_*$.
  
This complicated pattern can be understood and described analytically.
The analysis is quite parallel to what has been done earlier for the real 
walls.
In the limit $m \to 0$ and while $R(z)$ stays small as is the case in the
central wall region, the system (\ref{4sys}) is greatly simplified and
acquires the form
  \beq
 \label{urser}
\left\{ \begin{array}{l}
\rho' = (N-1)m\rho \sin \gamma \\
\gamma' = 2(N-1) m \cos \gamma
\end{array} \right.
  \eeq
It has the analytic solution
  \beq
  \label{solser}
\left\{ \begin{array}{l}
\rho(z) \ =\ \rho_0 \cosh^{1/2} [2m(N-1)z] \\
\sin \gamma(z) \ =\ \tanh [2m(N-1)z]
\end{array}  \right.
  \eeq
Indeed, the numerical solutions for $\rho(z)$ and $\gamma(z)$ follow the
formulae (\ref{solser}) rather closely up to the turning point at
$z \approx 20$. On the other hand, for the outer 
region  of the wall we are in the Higgs region. The fields $R(z)$ and 
$\beta(z)$ are frozen as is dictated by Eq.(\ref{freeze}), and it is the
equation system (\ref{rogam}) rather than Eq.(\ref{urser}) which describes
the wall dynamics. The change of the regime occurs when the trajectory 
(\ref{solser}) intercepts the integral trajectory of Eq.(\ref{rogam}).
For $N \geq 3$, the dependencies $\rho(z)$ and $\gamma(z)$ in the Higgs
phase can be found only numerically, but the trajectory in 
the ($\rho,\ \gamma$) plane is described analytically according to 
Eq.(\ref{intmot}). The interception occurs at

  \beq
  \label{cross}
\rho_\times \ =\ \frac{\rho_0}{\sqrt{\cos \frac \pi{2(N-1)}}}, \ \ \ 
\gamma_\times \ =\ \frac \pi{2(N-1)}
  \eeq
  
In the transitional region, we should fix instead $\rho(z)$ and $\alpha(z)
= \gamma(z)/2 + \pi/(2N)$ to their values in Eq.(\ref{cross}) after which the
system (\ref{4sys}) is reduced to
  \beq
  \label{trans}
R' \ =\ -2R^2 \ln [R^3 \rho_\times^{2(N-1)} ]
  \eeq
and $\beta$ is fixed at the value
 \beq
\beta_\times \ =\ -\frac{N-1}3 \left(\gamma_\times + \frac \pi N \right)
\ =\ \frac \pi {3N} - \frac \pi 2
   \eeq
After a trivial resclaling, the equation (\ref{trans}) just coincides with 
the BPS equation (\ref{fiN}) describing the real wall in pure supersymmetric 
gluodynamics. As was the case for the real walls, the transitional region, 
 the
region where the potential barrier between the Higgs phase 
$\phi = \chi^{-2(N-1)/3}$ and the symmetric phase $\phi = 0$ is penetrated,
is rather narrow. Its width in $z$ is of order $\rho_\times^{2(N-1)/3}
\propto m^{-(N-1)/3N}$. On the other hand, the characteristic width of the 
central and of the outer region is large $\propto 1/m$.

Note that, for $N=2$ when our wall is a superposition of two distant real 
walls,
its structure is simpler. A wide outer region is absent simply because the
values (\ref{cross}) where the transition occurs coincide in this case with
the asymmetric vacuum values. But the wide central region and the narrow
transitional region are still there.

The full energy of the configuration described coincides, of course, for small
masses with the BPS bound (\ref{epsc}). One can find out that the fractions
of the wall energy $f_{c.},\ f_{\rm trans.}$ and $f_{\rm out}$ carried, 
correspondingly, by the central region, the transitional region, and the 
outer   region are
\footnote{To avoid confusion, note that, by derivation, this result is valid
only for $N \geq 3$.}

  \beq
 \label{fract}
f_{c.} \ &=&\ \frac {\tan \frac{\pi}{2(N-1)}}{\tan \frac \pi N},\ \ \ 
 f_{\rm trans.} \ =\ \frac 1{N \sin \frac \pi N}
\left[ \frac {(N-1) \cos \frac \pi {2(N-1)}}{N \cos \frac \pi N} 
\right]^{N-1}\, , \nonumber \\ 
f_{\rm out} \ &=&\ 1 - f_{c.} - f_{\rm trans.} \hfill
  \eeq
For $N=3$, $f_{\rm c.} \approx$  .58, $ f_{\rm trans.} \approx$ .34, and
$f_{\rm out} \approx$ .08.

\section{Discussion}
\setcounter{equation}0 
 
Our main result is that, while the real BPS domain walls connecting 
the chirally symmetric and a chirally asymmetric vacua are present at
all masses, the complex BPS walls interpolating between different asymmetric
vacua exist only for small enough masses $m < m_*$, $m_*$ being
 given in Eq.(\ref{mcrit}).
  A kind of phase transition associated with the restructuring of the wall
spectrum occurs.
\footnote{Needless to say, it is not  a phase transition of a habitual
 thermodynamic variety. In particular, the vacuum energy is zero both
  below and above the phase transition point --- supersymmetry is never
  broken here. Hence $E_{vac} (m) \equiv 0$ is not singular at 
$m = m_*$.}
 It makes sense to
express the result in invariant terms and to trade $\Lambda_{SQCD}$ for an    
 invariant physical quantity such as the gluino condensate $\Sigma = |
<{\rm Tr}\ \lambda^2>| $ in a chirally asymmetric vacuum. Restoring the
dimensional factor $\Lambda_{SQCD}^{(2N+1)/N}$ in Eqs.(\ref{cond}, \ref{condk})
and combining it with Eq.(\ref{mcrit}),  we obtain 
   \beq
 \label{mSig}
m_*^{SU(3)} \ \approx\ .085\Sigma^{1/3} ,\ \ \ \  
m_{*}^{SU(4)} \ \approx \ 0.033 \Sigma^{1/3}
 \eeq
We cannot, however, claim that the {\it quantitative} estimates in 
Eq.(\ref{mSig}) are, indeed, quite correct. The matter is that the particular
values of the factors in Eq.(\ref{mSig}) are sensitive to the form of the 
kinetic term in the effective lagrangian which, in contrast to the
potential term, is not fixed rigidly by symmetry considerations. Just 
multiplying, say, the standard kinetic term for the $X$ superfield 
${\cal L}_{\rm kin}^X = 1/4 \int d^4\theta \bar X X$  by a numerical 
factor would
change the particular values (\ref{mSig}) of $m_*$ \cite{SVn}. 
The effective lagrangian may also involve complicated kinetic structures
with higher field derivatives $\propto (\partial \phi)^4$ etc.

The effective TVY lagrangian {\it implies} the presense of the chirally
invariant phase \cite{Kovner}. Recently, this conclusion has been criticized
\cite{Csaki,Schwetz}. In particular, it was shown in Ref.\cite{Csaki}
that the {\it assumption} that the spectrum of the chirally symmetric phase
in the supersymmetric gluodynamics involves the massless particles associated 
with the colorless composite 
field $\Phi$ in the effective lagrangian does not conform with the 't Hooft
 anomaly matching conditions of some global discrete anomalies.

However, this does not exclude the possibility that the chirally symmetric 
phase still exists, but its spectrum has nothing to do with that extracted
 from the naive  VY effective lagrangian with the standard kinetic term.
Indeed, for the dimension of the above mentioned terms with higher derivatives
of $\phi$ to be correct, they should involve some powers of $\phi$ in the
denominator (no new dimensionful constants can be introduced: that would spoil
the conformal properties of the lagrangian). The presense of such singular 
terms would modify the spectrum of the chirally symmetric phase completely
\cite{remark}. Actually, the dynamics of the symmetric phase is quite unclear
by now. This is an extremely interesting problem to solve but, at the moment,
 we do not have insights in this direction.

We want to emphasize, however, that though higher--derivative singular terms
may destroy completely a naive spectrum picture, they do not affect the 
conclusion on the {\it existence} of the symmetric phase. When the fields are
{\it static and homogeneous}, the form of the effective lagrangian is
extracted quite rigidly from the requirement that the conformal and chiral 
anomaly of the original theory are reproduced correctly. Also, the dynamics
of the domain walls depend largely on the region {\it between} the vacua
where the extra terms with higher derivatives are not singular. Thereby, they
should not modify the structure of the walls revealed in Refs.\cite{my,SV,SVn}
and in this paper. At least, this is our guess and hope.
In particular, a {\it qualitative} conclusion on the absence of a smooth
transition between the small mass region and the large mass region  in 
supersymmetric QCD {\it is} correct.

In this paper, we studied only the BPS solutions, but,
as it was done earlier for $SU(2)$ \cite{SVn}, one may and actually should
   study also non--BPS walls, the field
configurations which satisfy the equations of motion but not the first--order
BPS equations. This work is now in progress \cite{prep}. Our preliminary 
results display that the picture is roughly the same as for $SU(2)$: in some
range of masses $m_* < m \leq m_{**}$, a non-BPS complex domain wall 
presenting a local minimum of the energy functional exists. There are also
sphaleron wall solutions. For $m > m_{**}$,
the complex walls disappear  altogether from the spectrum.

\vspace{.5cm}

{\bf Acknowledgments}: \hspace{0.2cm} 
 I acknowledge illuminating discussions with A. Kovner and A.I. Veselov and 
warm hospitality  extended to me at Saclay where this work was finished.
This work was supported in part  by the RFBR--INTAS grants 93--0283 and
 94--2851, by the RFFI grant 97--02--16131, 
by the U.S. Civilian Research and Development Foundation under award 
\# RP2--132, and by the Schweizerishcher National Fonds grant \# 7SUPJ048716.

\end{document}